\documentclass[prb,aps,twocolumn,nofootinbib,10pt, showpacs]{revtex4-1}
\usepackage{amssymb}
\usepackage{appendix}
\usepackage{times}
\usepackage{graphicx}
\usepackage{subeqnarray}
\newcommand{\ket}[1]{|#1\rangle}
\newcommand{\bra}[1]{\langle #1|}
\usepackage{color}

\begin{document}

\title {Robust Dual Topological Character with Spin-Valley Polarization in a Monolayer of the Dirac Semimetal Na$_3$Bi}

\author{Chengwang Niu$^1$}
\email{c.niu@fz-juelich.de}
\author{Patrick M. Buhl$^1$}
\author{Gustav Bihlmayer$^1$}
\author{Daniel Wortmann$^1$} 
\author{Ying Dai$^2$}
\author{Stefan Bl\"{u}gel$^1$}
\author{Yuriy Mokrousov$^1$}
\affiliation
{$^1$Peter Gr\"{u}nberg Institut and Institute for Advanced Simulation, Forschungszentrum J\"{u}lich and JARA, 52425 J\"{u}lich, Germany
\\$^2$School of Physics, State Key Laboratory of Crystal Materials, Shandong University, 250100 Jinan, People's Republic of China}

\begin{abstract}
Topological materials with both insulating and semimetal phases can be protected by crystalline (e.g. mirror) symmetry. The insulating phase, called topological crystalline insulator (TCI), has been intensively investigated and observed in three-dimensional materials. However, the predicted two-dimensional (2D) materials with TCI phase are explored much less than 3D TCIs and 2D topological insulator, while so far considered 2D TCIs almost exclusively possess a square lattice structure with the mirror Chern number $\mathcal C_{M} =-2$. Here, we predict theoretically that hexagonal monolayer of Dirac semimetal Na$_3$Bi is a 2D TCI with a mirror Chern number $\mathcal C_{M} =-1$. The large nontrivial gap of 0.31 eV is tunable and can be made much larger via strain engineering while the topological phases are robust against strain, indicating a high possibility for room-temperature observation of quantized conductance. In addition, a nonzero spin Chern number $\mathcal C_{S} =-1$ is obtained, indicating the coexistence of 2D topological insulator and 2D TCI, i.e. the dual topological character. Remarkably, a spin-valley polarization is revealed in Na$_3$Bi monolayer due to the breaking of crystal inversion symmetry. The dual topological character is further explicitly confirmed via unusual edge states' behavior under corresponding symmetry breaking. 
\end{abstract}

\maketitle
\date{\today}

The discovery of topological insulator (TI)~\cite{Hasan,Qi1} has triggered an explosion of novel topologically nontrivial phases, such as the topological crystalline insulator (TCI), for which the role of the time-reversal symmetry is replaced by the crystal (mirror) symmetry~\cite{Futci,Hsieh,Ando}. The hallmark of a TCI, similar to a TI, is the presence of gapless surface/edge states with Dirac points inside of the insulating bulk energy gap. In presence of crystal mirror symmetry, the coexistence of TI and TCI phases has been predicted in three dimensions (3D) for Bi$_{1-x}$Sb$_x$~\cite{Teo} and Bi chalcogenides~\cite{Rauch,Weber,Eschbach}, and thus they exhibit a dual topological character (DTC). Recently, unusual topological surface states for a 3D DTC system have been observed experimentally~\cite{Weber,Eschbach}. In the 2D case, graphene maybe a prototypical example of DTC~\cite{Kane,Liu}. However, the extremely small band gap of graphene makes it very difficult to verify the DTC in this material experimentally~\cite{Yao}. To date, the 2D TIs are identified experimentally in HgTe/CdTe~\cite{konig} and InAs/GaSb~\cite{Knez} quantum wells at low temperatures, and a lot of 2D TIs  with giant band gaps have been predicted to exist as a result of substrate interaction effect~\cite{zhoum}, chemical functionalization~\cite{Xuy,Song,niubi,LiL,Crisostomo}, or global structure optimization~\cite{luo}. In many cases the complex structures and the lack of mirror symmetry in such materials forbid the formation of a 2D TCI phase. On the other hand, 2D TCIs are so far limited to theoretical predictions that are mainly restricted to SnTe multilayers~\cite{Liu}, (Sn/Pb)(Se/Te) monolayers~\cite{Wrasse,Liu2,niutci}, Tl(S/Se) monolayers~\cite{niutlse}, and SnTe/NaCl quantum wells~\cite{niu2d} with mirror Chern number $\mathcal C_{M}=-2$. The even number of band inversions leads to a vanishing $\mathbb{Z}_2$  invariant. Therefore, 2D TCIs with $\mathcal C_{M}=-2$ cannot be 2D TIs protected by the time reversal symmetry. Thus, for further investigation and applications of DTC in two dimensions, it is essential to extend the domain of candidate 2D TIs and TCIs both with respect to topological manifestations (i.e. different $\mathcal C_{M}$~\cite{Takahashi}) and material realisation.

For both 2D TIs and 2D TCIs, the spin-orbit coupling (SOC) is known to play a vital role. In addition, the SOC together with inversion symmetry breaking can lead to coupled spin and valley physics, in which the new degree of freedom offers a promising route to the eventual realization of valleytronic devices~\cite{Rycerz,xiao}. The spin-valley polarization has been observed experimentally in MoS$_2$ monolayer~\cite{Mak}, which is  a topologically trivial insulator. Therefore, a natural question arises as to whether the spin-valley polarization in nontrivial insulators, such as 2D TIs and 2D TCIs, is possible. Recently, thin films of the Dirac semimetal Na$_3$Bi~\cite{Wang1,Liuz} have been fabricated by molecular beam epitaxy~\cite{Hellerstedt} and, therefore, in the present study, we take Na$_3$Bi as an example and propose the realization of the 2D DTC in a monolayer of Na$_3$Bi with a band gap of 0.31 eV, which is well above the energy scale of room temperature. The calculated spin Chern number $\mathcal C_{S} =-1$ and mirror Chern number $\mathcal C_{M} =-1$ confirm the 2D DTC phase directly. In addition, the spin-valley polarization due to the lack of the spatial inversion symmetry is investigated.

The density functional calculations are performed using the generalized gradient approximation (GGA) of Perdew-Burke-Ernzerhof (PBE)~\cite{Perdew} for the exchange correlation potential as implemented in the \texttt{FLEUR} code~\cite{fleur} as well as in the Vienna {\it ab-initio} simulation package (VASP)~\cite{Kresse,Kresse1}. A 20 {\AA} thick vacuum layer is used to avoid interactions between nearest slabs for VASP while the film calculations are carried out with the film version of the \texttt{FLEUR} code~\cite{Krakauer}. SOC is included in the calculations self-consistently. The maximally localized Wannier functions (MLWFs) are constructed using the {\tt wannier90} code in conjunction with the \texttt{FLEUR} package.~\cite{Mostofi,Freimuth} 

\begin{figure}[!t]
\centering
\includegraphics{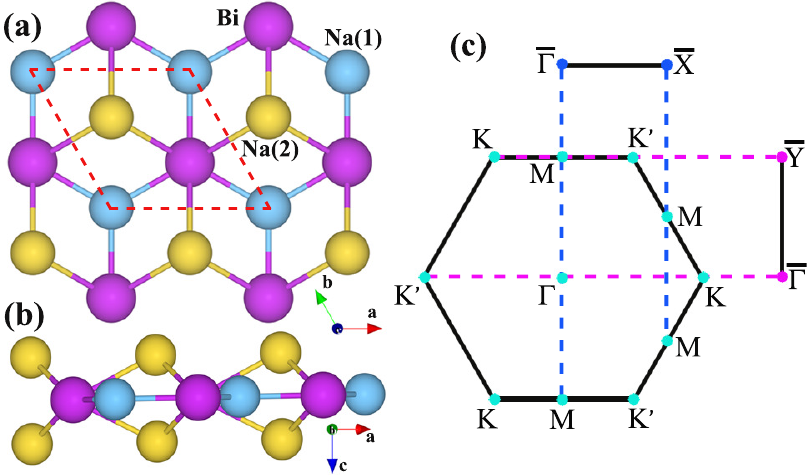}
\caption{ 
(a) Top and (b) side view of honeycomb Na$_3$Bi monolayer, where the unit cell is indicated by the dashed lines. (c) The Brillouin zone of 2D Na$_3$Bi monolayer and the projected 1D Brillouin zones.}
\label{structure}
\end{figure}

Bulk Na$_3$Bi in hexagonal $P6_3/mmc$ structure is a three-dimensional (3D) counterpart of graphene, which hosts 3D Dirac points in its electronic structure and is called a topological Dirac semimetal~\cite{Wang1,Liuz}. The bulk crystal structure consists of stacked triple layers along the $z$-direction. Each triple layer has four atoms which are one Bi in Wyckoff 2c position, one Na(1) in 2b position, and two Na(2) in 4f position. In Figs.~\ref{structure}(a) and (b) the side and top view of the Na$_3$Bi triple layer are presented, with Bi and Na atoms forming the honeycomb lattice. Unlike in the bulk material, the inversion symmetry is broken in  a Na$_3$Bi triple layer but the mirror symmetry $z\rightarrow -z$ is preserved,  in exact analogy to a MoS$_2$ monolayer~\cite{Splendiani}. Hereafter, we call such a triple layer a Na$_3$Bi monolayer. To check its energetic stability, the formation energy is calculated by $E_f = E_{\rm Na_3Bi} - 3\mu_{\rm Na}-\mu_{\rm Bi}$, where E$_{\rm Na_3Bi}$ is the total energy of the Na$_3$Bi monolayer,  $\mu_{\rm Na}$ and $\mu_{\rm Bi}$ are the chemical potentials of Na and Bi atoms, respectively. The calculated formation energy of $-0.78$ eV indicates that the Na$_3$Bi monolayer is energetically stable .

\begin{figure}
\centering
\includegraphics{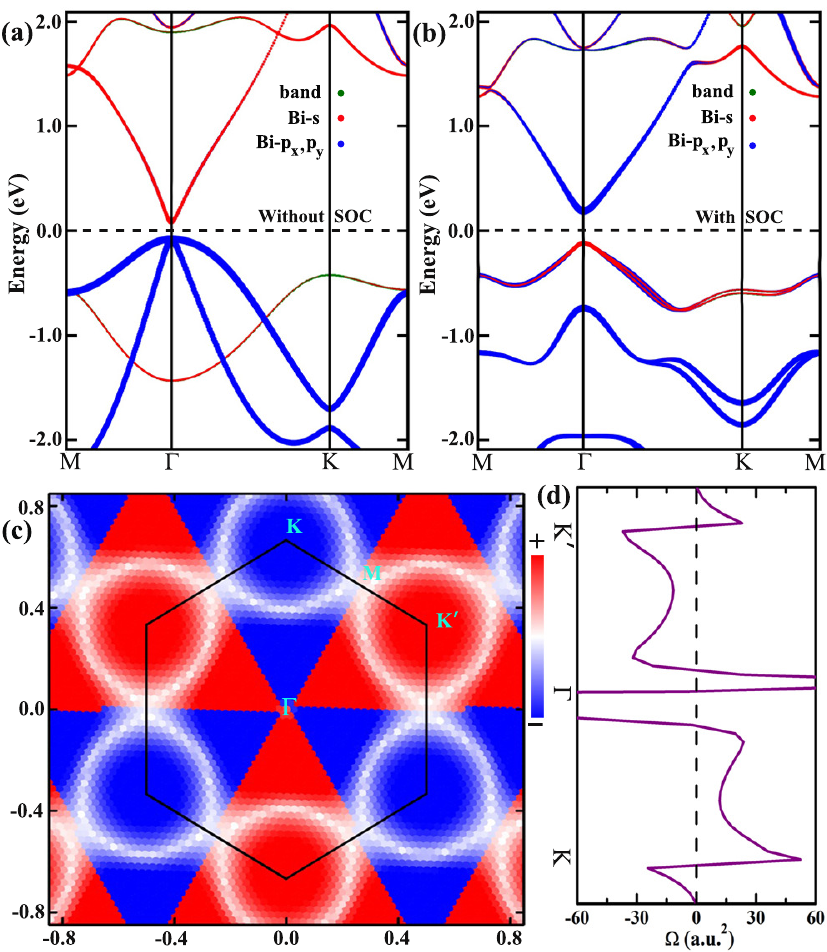}
\caption{Orbitally-resolved band structures for Na$_3$Bi monolayer (a) without and (b) with SOC, weighted with the contribution of Bi-s and Bi-p$_x$,p$_y$ states. The Fermi level is indicated with a dashed line. (c) Momentum-resolved polarization of spin perpendicular to the mirror plane for the highest occupied band. (d) Berry curvature distribution of the highest occupied bands in the $K-\Gamma-K^{\prime}$ direction.}
\label{band}
\end{figure}

Figures~\ref{band}(a) and (b) present the orbitally resolved band structures of Na$_3$Bi monolayer without and with SOC, respectively, that deliver preliminary insight into topological properties of the system. Due to the presence of time-reversal symmetry, the bands at valleys K and K$^\prime$ are energetically degenerate, and thus we only show the dispersion around K. In the absence of SOC, Bi-p$_x$ and Bi-p$_y$ orbitals contribute to the valence band maximum (VBM) while the conduction band minimum (CBM) is dominated by Bi-s orbitals with a direct band gap of 0.16 eV. Switching on SOC leads to an inversion of the VBM and the CBM, and an s-p band inversion occurs at $\Gamma$ point. The insulating character is preserved with a band gap of 0.31 eV, indicating the feasibility of experimental observation of the 2D topological properties of this material at room temperature. To further confirm our results, the band structure is checked by using the more sophisticated Heyd-Scuseria-Ernzerhof hybrid functional method (HSE06)~\cite{Heyd}. Similar to 1-T$^\prime$ MoS$_2$~\cite{Qian}, the nontrivial phase has a larger band gap ($\sim$ 0.4 eV) when SOC is included.
 
The existence of the mirror symmetry $z\rightarrow -z$ (see Figs.~\ref{structure}(a) and (b)) for Na$_3$Bi monolayer promises the possibility of realizing the TCI that is characterized by the so-called mirror Chern number~\cite{Futci,Teo}, which is defined as $\mathcal C_{M}= (\mathcal C_{+i}-\mathcal C_{-i})/2$, where $\mathcal C_{+i}$ and $\mathcal C_{-i}$ are the Chern numbers for mirror eigenvalues $+i$ and $-i$,~\cite{yao2004}
\begin{eqnarray}
	\mathcal C_{\pm i}=\frac{1}{2\pi} \sum_{n<E_F}\int_{BZ} \Omega_{\pm i}({\bf k}) d^{2}k~,
\end{eqnarray}
and $\Omega_{\pm i}({\bf k})$ is the Berry curvature of all occupied bands constructed from respective mirror projected states in the mirror plane, calculated according to
\begin{eqnarray}
	\Omega({\bf k})=-2{\rm Im}\sum_{m\ne n}\frac{\bra{\psi_{n{\bf k}}}\upsilon_x\ket{\psi_{m{\bf k}}}
		\bra{\psi_{m{\bf k}}}\upsilon_y\ket{\psi_{n{\bf k}}}}{(\varepsilon_{m{\bf k}}-\varepsilon_{n{\bf k}})^2},
\end{eqnarray}
where $m, n$ are band indices, $\psi_{m/n{\bf k}}$ and $\varepsilon_{m/n{\bf k}}$ are the corresponding wavefunctions and eigenenergies of band $m/n$, respectively, and $\upsilon_{x/y}$ are the velocity operators. The MLWFs are constructed to calculate the Berry curvature efficiently. With $z\rightarrow -z$ mirror symmetry, the calculated Chern numbers are respectively $\mathcal C_{\pm i}=\mp 1$, leading to a mirror Chern number $\mathcal C_{M}= -1$. This indicates that the Na$_3$Bi monolayer is a 2D TCI. Interestingly, the $\mathcal C_{M}= -1$ case we consider here is topologically distinct from the previously reported 2D TCIs, such as (Sn/Pb)Te~\cite{Liu,Liu2,niutci} and Tl(S/Se)~\cite{niutlse} with $\mathcal C_{M}= -2$. 

Here, in Na$_3$Bi monolayer, band inversion occurs at $\Gamma$ point, i.e., we acquire an odd number of band inversions. To identify the relationship between the 2D TI and the odd number of band inversions in Na$_3$Bi monolayer, we calculate the spin Chern number $\mathcal C_{S}$~\cite{Yang,Prodan,zhanghb} which can be directly related to the $\mathbb{Z}_2$ topological invariant of the system. $\mathcal C_{S}$ provides equivalent characterization to $\mathbb{Z}_2$ number in that for time-reversal symmetric and inversion symmetric systems the even values of $\mathcal C_{S}$ correspond to a topologically trivial insulator state, while odd values of $\mathcal C_{S}$ indicate the emergence of a TI phase~\cite{Yang,Prodan,zhanghb}. The $\mathcal C_{S}$ is given by the difference of the Chern numbers for the spin-up ($\mathcal C_{+}$) and spin-down ($\mathcal C_{-}$) projected manifolds, $\mathcal C_{S}= (\mathcal C_{+}-\mathcal C_{-})/2$~\cite{Yang}. The $\sigma_z$ matrix, $\bra{\phi_{m{\bf k}}}\sigma_z\ket{\phi_{n{\bf k}}}$, is constructed and diagonalized to distinguish the spin-up and spin-down manifolds~\cite{zhanghb}. The Chern number for each spin manifold is $\mathcal C_{+} = -1$ and $\mathcal C_{-} = 1$, yielding the spin Chern number $\mathcal C_{S} = -1$. This clearly demonstrates the 2D TI nature of the Na$_3$Bi monolayer. Therefore, Na$_3$Bi monolayer exhibits the DTC with respect to the 2D TI and 2D TCI phases. 

\begin{figure}
\centering
\includegraphics{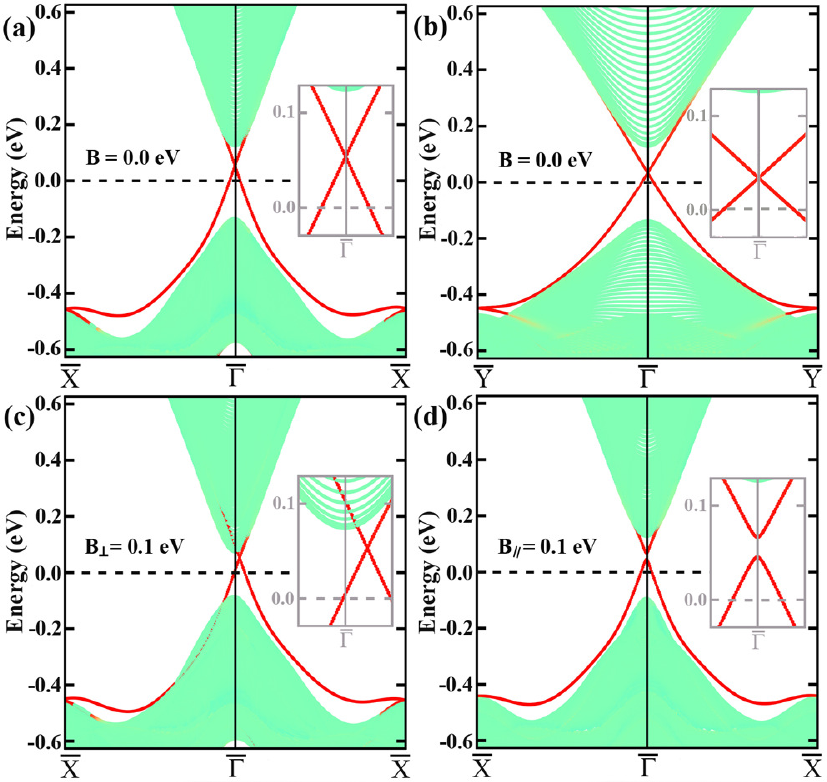}
\caption{Localization-resolved edge states of Na$_3$Bi monolayer for different configurations. (a) Bi-Na(1) termination without a magnetic field, (b) Bi-Na(1)-Na(2) termination without a magnetic field, (c) Bi-Na(1) termination a magnetic field perpendicular to the mirror plane, and (d) Bi-Na(1) termination with a magnetic field within the mirror plane. Insets show the corresponding zoom-in at the $\bar\Gamma$ point. Color from light green to red represents the weight of atoms located from middle to one edge of the ribbon structures.
}
\label{edge}
\end{figure}

To further confirm the DTC, we investigate the edge states of 1D nanoribbons of Na$_3$Bi monolayer. The nonzero $\mathcal C_{M}$ and/or $\mathcal C_{S}$ should support the gapless edge states bridging the conduction and valence bands, which exhibit a band gap when both the time-reversal and mirror symmetries are broken. Based on a description in terms of MLWFs of the Na$_3$Bi monolayer, the tight-binding Hamiltonians of nanoribbons along two different directions with a width of 60 unit cells are constructed. The calculated band structures on one side of the ribbons are presented in Figs.~\ref{edge}(a) and (b), respectively. One can clearly see a pair of gapless edge states in the 2D gap. The Dirac point around $\bar\Gamma$ is a clear consequence of the non-trivial topological character of the system.

Generally, time-reversal symmetry breaking generates a gap in the surface/edge states of TIs~\cite{Hasan,Qi1} while mirror symmetry breaking is indispensable for the formation of a band gap in the surface/edge states of TCIs~\cite{Futci}. One way to destroy these symmetries is to introduce the magnetism in the system. To mimic a magnetic environment, we compute the matrix elements of the Pauli matrices $\sigma_\alpha$ ($\alpha= x,y,z$) in the basis of MLWFs, which allows us to consider the effect of an exchange field applied along different directions. For an exchange field perpendicular to the mirror plane, $H_{mag}=B_{\bot}\cdot\sigma_z$,  the time-reversal symmetry is broken while the mirror symmetry is maintained. In this case, as shown in Fig.~\ref{edge}(c) for the Bi-Na(1) termination, the Dirac point moves slightly away from the $\bar\Gamma$ point, while
a band gap does not open as a consequence of 2D TCI phase's survival. If the exchange field, on the other hand, is in the plane, $H_{mag}=B_{\parallel}\cdot\sigma_{x}$, both time-reversal and mirror symmetries are broken and the edge states become gapped (Fig.~\ref{edge}(d)). This behavior is reminiscent of that in Bi$_2$(Se/Te)$_3$~\cite{Wray,Rauch}, but with different directions of an exchange field owing to a different sense of the mirror sysmmetry in these two compounds (B$_z$ in the 3D TI corresponds to B$_{\parallel}$ in the 2D TI)~\cite{Wray,Rauch}.

Exposing honeycomb lattices to inversion symmetry breaking provides a new, so-called valley, tunable degree of freedom in addition to spin and charge. The valley degree of freedom is receiving considerable attention these days due to potential application in valleytronics~\cite{xiao,Mak}. To demonstrate the effect of the inversion symmetry breaking, we focus now on the spin-valley coupling of the Na$_3$Bi monolayer by considering the spin-polarization of occupied states in reciprocal space. Since the in-plane components of the spin polarization are vanishing due to the presence of mirror symmetry, in Fig.~\ref{band}(c) we plot the momentum-resolved out-of-plane spin polarization of the highest occupied band. It is clearly visible that the highest occupied state exhibits the spin polarization which is of opposite sign at valleys $K$ and $K^{\prime}$. We further inspect the Berry curvature of the highest occupied band along the $K-\Gamma-K^{\prime}$ path, plotted in Fig.~\ref{band}(d). An odd behavior of the Berry curvature  with respect to the valley agrees with the symmetry analysis in terms of time- and structural-inversion symmetry, similarly to the well-known case of spin-valley coupling in MoS$_2$ monolayer~\cite{Cao}. Valley polarization that is coupled with spin will suppresses spin and valley relaxation and is promising to prepare the information carriers for the next-generation electronic and optoelectronic devices~\cite{Mak}.

\begin{figure*}[!t]
\includegraphics{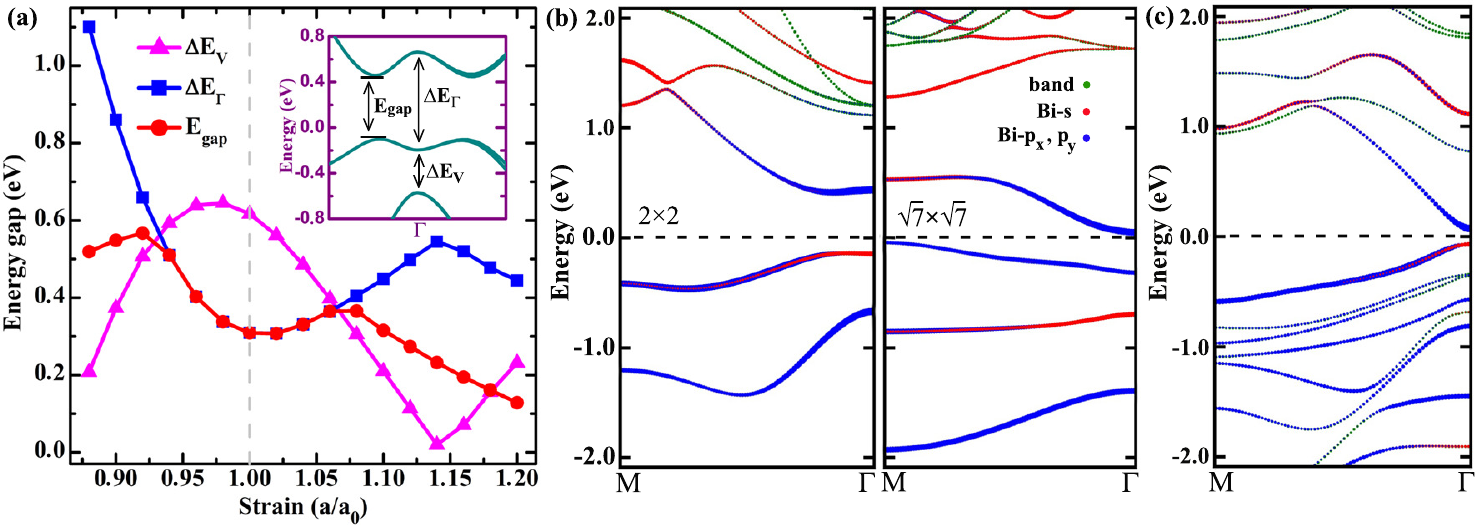}
\caption{(a) Variation of the energy gaps for Na$_3$Bi monolayer as a function of strain. The gaps ($\Delta$E$_{\Gamma}$ and E$_{gap}$ in inset) between the valence and conduction bands remain while a strain-induced band inversion between the valence bands ($\Delta$E$_V$ in inset) occurs. (b) Orbitally-resolved band structures for Na$_3$Bi monolayer sandwiched between a substrate of (b) BN with $2\times2$ and $\sqrt{7}\times\sqrt{7}$ supercells and (c) Na$_3$Sb, weighted with the contribution of Bi-s and Bi-p$_x$,p$_y$ states. The DTC remains intact with corresponding substrates.}
\label{strain}
\end{figure*}

Having established our material's DTC, accompanied by a band gap of 0.31 eV  (which is large enough for practical applications at room temperature), we finally test its stability. The phonon spectrum calculation shows imaginary frequencies around the M point, but not at the $\Gamma$ point, indicating that the band inversion at this point is robust. We demonstrate this by investigating the band inversion under various strains as well as substrates~\cite{YangK}. The magnitude of strain is described by the ratio a/a$_0$, where "a$_0$" and "a" denote the lattice parameters of the unstrained (5.31 {\AA}) and strained systems, respectively. The results of the calculations shown in Fig.~\ref{strain}, indicate that the band gap of Na$_3$Bi monolayer can be significantly enhanced upon straining, similarly to the results reported previously~\cite{niutlse,niubi}. There is no band-gap closing-reopening process between the valence and conduction bands ($\Delta$E$_{\Gamma}$ and E$_{gap}$) in a large range of strain from $-12$\% to $20$\%. The persistent band inversion indicates that the topological character of the system is robust against the substrate-imposed strain, which is important for further experimental investigations and device applications due to the fact that a freestanding film is usually hard to grow. Figure~\ref{strain}(b) shows the calculated band structures of quantum well structures, that retain the mirror symmetry, with Na$_3$Bi monolayer sandwiched between $2\times2$ and $\sqrt{7}\times\sqrt{7}$ BN monolayers. As we can see, the energy gap at the M point changes much stronger than at $\Gamma$ with the inverted band gap at the $\Gamma$ point surviving in both cases (although a further band inversion occurs between valence bands for the $\sqrt{7}\times\sqrt{7}$ case, see also $\Delta$E$_V$ of Fig.~\ref{strain}(a)), which proves that DTC is preserved with respective substrate-induced strain of $-6.2$\% and $19.7$\%. A weak interlayer interaction is expected between BN and Na$_3$Bi because of the large relaxed interlayer distances $\sim$4.0 \AA. We then consider a strong interlayer interaction via the topologically trivial substrate Na$_3$Sb that has the same crystal structure as Na$_3$Bi but creates $-2$\% epitaxial strain. As shown in Fig.~\ref{strain}(c), the band inversion remains at the $\Gamma$ point.
 
In summary, based on first-principles calculations, we revealed that the 2D TI and 2D TCI phases can coexist in Na$_3$Bi monolayer with a large band gap of 0.31 eV. Nonzero spin Chern number and mirror Chern number, as well as nontrivial topological edge states confirm the dual topological character clearly.  As the mirror symmetry is perserved for an out-of-plane exchange field, the gapless edge states survive but move away from the time-reversal invariant momenta, while a gap opens for in-plane exchange fields. At last, strain engineering shows that the dual topological character in Na$_3$Bi monolayer is robust in a large range of strain in which the nontrivial band gap can be tuned efficiently.

This work was supported by the Priority Program 1666 of the German Research Foundation (DFG) and the Virtual Institute for Topological Insulators (VITI). We acknowledge computing time on the supercomputers JUQUEEN and JURECA at J\"{u}lich Supercomputing Centre and JARA-HPC of RWTH Aachen University.

\end{document}